\documentclass[aps,prl,preprint,tightenlines,superscriptaddress,showpacs,byrevtex,subfigure]{revtex4}

\usepackage{graphicx}
\usepackage{color}

\begin{document}
\begin{flushright}
Belle Preprint 2007-47\\
KEK  Preprint 2007-57
\end{flushright}

\title{ \quad\\[0.5cm] 
Search for Lepton Flavor
Violating $\tau$ 
{Decays}\\
into Three Leptons
}

\affiliation{Budker Institute of Nuclear Physics, Novosibirsk, Russia}
\affiliation{Chiba University, Chiba, Japan}
\affiliation{University of Cincinnati, Cincinnati, OH, USA}
\affiliation{The Graduate University for Advanced Studies, Hayama, Japan}
\affiliation{Hanyang University, Seoul, South Korea}
\affiliation{University of Hawaii, Honolulu, HI, USA}
\affiliation{High Energy Accelerator Research Organization (KEK), Tsukuba, Japan}
\affiliation{Institute of High Energy Physics, Chinese Academy of Sciences, Beijing, PR China}
\affiliation{Institute for High Energy Physics, Protvino, Russia}
\affiliation{Institute of High Energy Physics, Vienna, Austria}
\affiliation{Institute for Theoretical and Experimental Physics, Moscow, Russia}
\affiliation{J. Stefan Institute, Ljubljana, Slovenia}
\affiliation{Kanagawa University, Yokohama, Japan}
\affiliation{Korea University, Seoul, South Korea}
\affiliation{Kyungpook National University, Taegu, South Korea}
\affiliation{\'Ecole Polytechnique F\'ed\'erale de Lausanne, EPFL, Lausanne, Switzerland}
\affiliation{Faculty of Mathematics and Physics, University of Ljubljana, Ljubljana, Slovenia}
\affiliation{University of Maribor, Maribor, Slovenia}
\affiliation{University of Melbourne, Victoria, Australia}
\affiliation{Nagoya University, Nagoya, Japan}
\affiliation{Nara Women's University, Nara, Japan}
\affiliation{National Central University, Chung-li, Taiwan}
\affiliation{National United University, Miao Li, Taiwan}
\affiliation{Department of Physics, National Taiwan University, Taipei, Taiwan}
\affiliation{H. Niewodniczanski Institute of Nuclear Physics, Krakow, Poland}
\affiliation{Nippon Dental University, Niigata, Japan}
\affiliation{Niigata University, Niigata, Japan}
\affiliation{University of Nova Gorica, Nova Gorica, Slovenia}
\affiliation{Osaka City University, Osaka, Japan}
\affiliation{Osaka University, Osaka, Japan}
\affiliation{Panjab University, Chandigarh, India}
\affiliation{Saga University, Saga, Japan}
\affiliation{University of Science and Technology of China, Hefei, PR China}
\affiliation{Seoul National University, Seoul, South Korea}
\affiliation{Sungkyunkwan University, Suwon, South Korea}
\affiliation{University of Sydney, Sydney, NSW, Australia}
\affiliation{Toho University, Funabashi, Japan}
\affiliation{Tohoku Gakuin University, Tagajo, Japan}
\affiliation{Tohoku University, Sendai, Japan}
\affiliation{Department of Physics, University of Tokyo, Tokyo, Japan}
\affiliation{Tokyo Metropolitan University, Tokyo, Japan}
\affiliation{Tokyo University of Agriculture and Technology, Tokyo, Japan}
\affiliation{Virginia Polytechnic Institute and State University, Blacksburg, VA, USA}
\affiliation{Yonsei University, Seoul, South Korea}
\author{Y.~Miyazaki} 
\affiliation{Nagoya University, Nagoya, Japan}
\author{I.~Adachi} 
\affiliation{High Energy Accelerator Research Organization (KEK), Tsukuba, Japan}
\author{H.~Aihara} 
\affiliation{Department of Physics, University of Tokyo, Tokyo, Japan}
\author{K.~Arinstein} 
\affiliation{Budker Institute of Nuclear Physics, Novosibirsk, Russia}
\author{V.~Aulchenko} 
\affiliation{Budker Institute of Nuclear Physics, Novosibirsk, Russia}
\author{T.~Aushev} 
\affiliation{\'Ecole Polytechnique F\'ed\'erale de Lausanne, EPFL, Lausanne, Switzerland}
\affiliation{Institute for Theoretical and Experimental Physics, Moscow, Russia}
\author{A.~M.~Bakich} 
\affiliation{University of Sydney, Sydney, NSW, Australia}
\author{V.~Balagura} 
\affiliation{Institute for Theoretical and Experimental Physics, Moscow, Russia}
\author{E.~Barberio} 
\affiliation{University of Melbourne, Victoria, Australia}
\author{A.~Bay} 
\affiliation{\'Ecole Polytechnique F\'ed\'erale de Lausanne, EPFL, Lausanne, Switzerland}
% ...
\collaboration{The Belle Collaboration}
\noaffiliation

\begin{abstract}
We search for lepton-flavor-violating $\tau$ decays 
into  three leptons (electron or muon) 
using
535 fb$^{-1}$ of data collected 
with the Belle detector at the 
KEKB asymmetric-energy $e^+e^-$ collider. 
No evidence for these decays is 
{observed, and}  
we set 90\% confidence level 
{upper limits on the branching fractions of
$(2.0\!-\!4.1)\times 10^{-8}$.}
These results improve {upon} our 
previously published upper limits by factors {of} 
{4.9 to 10.}
\end{abstract}
\maketitle
 \section{Introduction}

\begin{figure}
\begin{center}
       \resizebox{0.4\textwidth}{0.4\textwidth}{\includegraphics
        {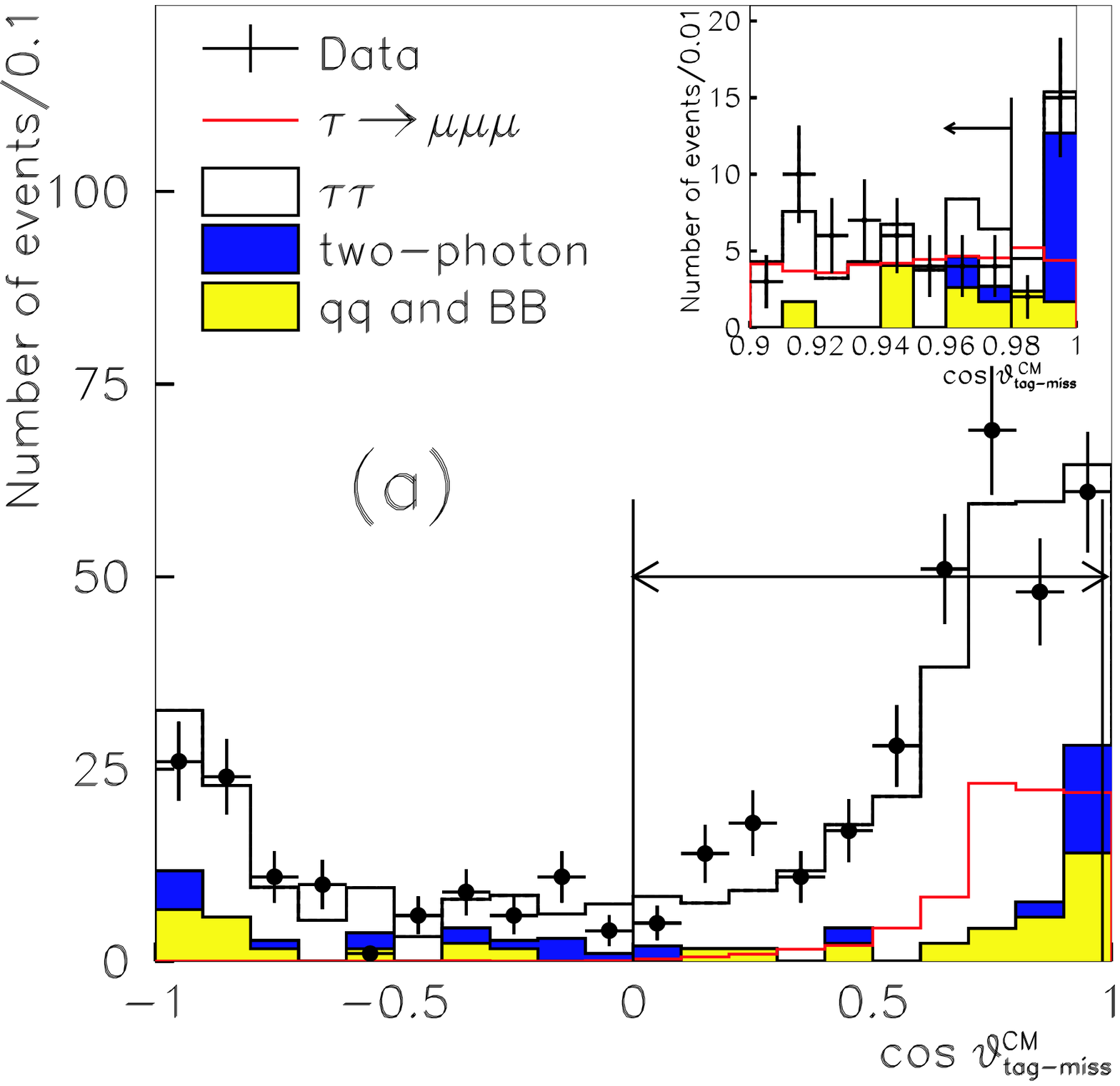}}
       \resizebox{0.4\textwidth}{0.4\textwidth}{\includegraphics
 {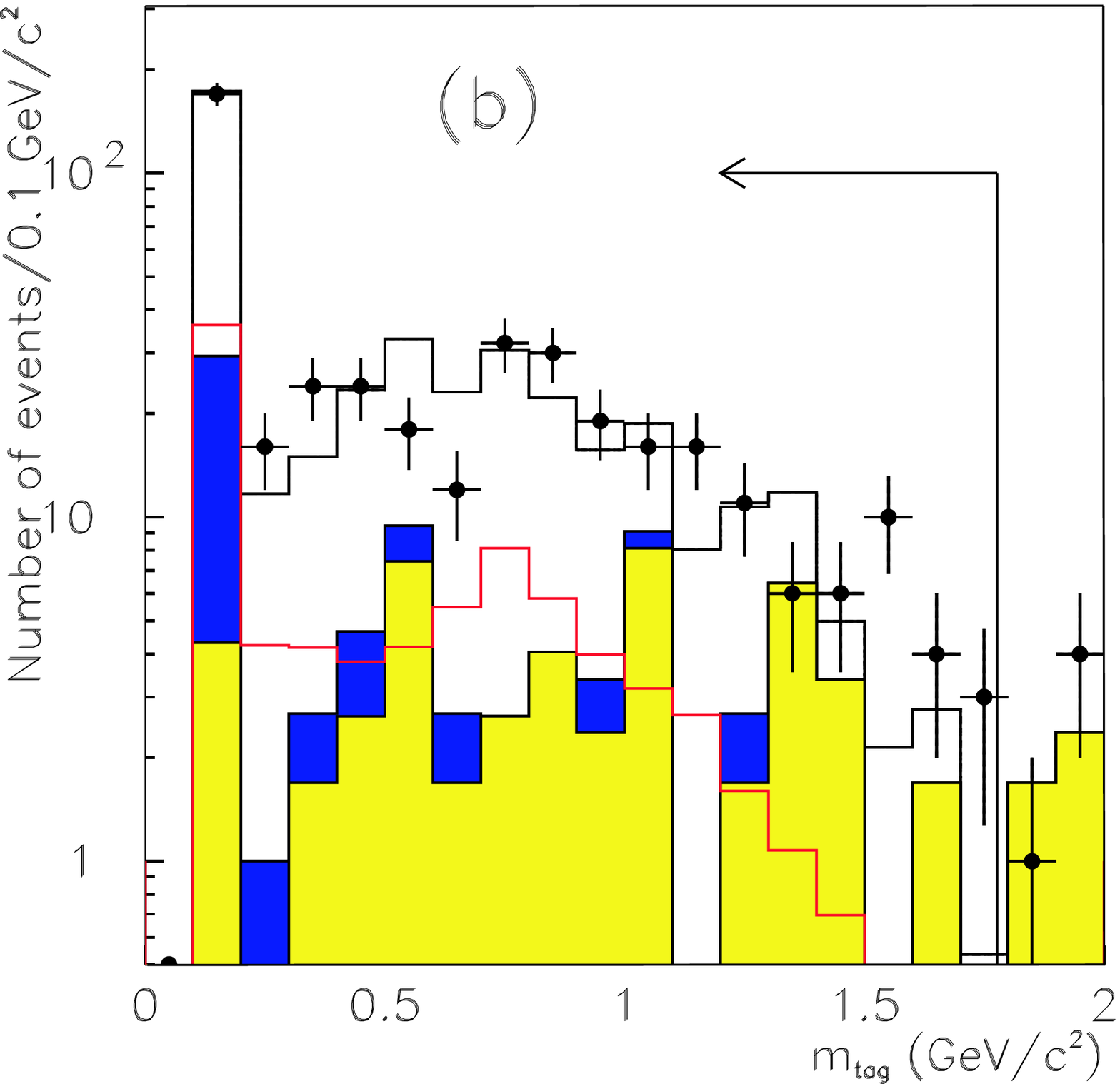}}
 \vspace*{-0.5cm}
 \caption{
 {Kinematic distributions used in the event selection:
 (a) the cosine of the opening angle between a charged track on the 
 tag side and
 missing particles in the CM system ($\cos \theta_{\rm tag-miss}^{\rm CM}$);
 (b) the reconstructed mass on the tag side using a charged track and photons
 after $E^{\rm CM}_{\rm vis}$ and $T$ event selection.
{The signal MC ($\tau^-\to\mu^-\mu^+\mu^-$)
 distributions with arbitrary normalization  are shown for
 comparison; 
 the background MC
 distributions are normalized to the data luminosity.}
 {Selected regions are indicated
 by {the}
 arrows from the marked cut {boundaries.}}}
}
\label{fig:cut_fig}
\end{center}
\end{figure}

{Lepton flavor violation (LFV)
{appears}
in {various} extensions of the Standard Model (SM).
{In particular, {lepton-flavor-violating} decays
$\tau^-\to\ell^-\ell^+\ell^-$
(where $\ell = e$ or $\mu$ )\footnotemark[2]
are discussed
{in 
various supersymmetry models~\cite{cite:susy1,
cite:susy2,cite:susy3,cite:susy4,cite:susy5,
cite:susy6,cite:susy7},
models with
{little Higgs~\cite{cite:littlehiggs1,cite:littlehiggs2},}
{{left-right symmetric models~\cite{cite:leftright}},}
as well as models with
heavy singlet Dirac neutrinos~\cite{cite:amon}
and
very light pseudoscalar bosons~\cite{cite:pseudo}.}
Some of these models with certain combinations of parameters 
predict that the branching fractions 
for $\tau^-\to\ell^-\ell^+\ell^-$ 
can be as high as $10^{-7}$, 
which is already accessible 
{in} high-statistics  $B$ factory experiments.
{Searches for LFV $\tau$ decays into three leptons have
a long history~\cite{13} 
{beginning from} the pioneering experiment
of MARKII~\cite{14}.
{In 
previous 
analyses,}}
both Belle and BaBar reached 
90\% confidence level (C.L.) upper limits 
on the branching fractions 
{at the  $10^{-7}$ level}~\cite{cite:3l_belle, cite:3l_babar}, based on 
about 90 fb${}^{-1}$ of data.
{The BaBar collaboration
has
recently
used 
376 fb${}^{-1}$ of data
to obtain 90\% C.L. 
upper limits 
in the range (3.7$-$8.0)~$\times~10^{-8}$~\cite{cite:3l_babar2}.}
Here, we update 
{our} previous results 
with {a} much larger data set (535 fb$^{-1}$) collected 
with the Belle detector 
{at the KEKB  
asymmetric-energy 
$e^+e^-$ 
collider~\cite{kekb},}
{taken at} 
the $\Upsilon(4S)$ resonance and 60 MeV below it.}

\footnotetext[2]{{Throughout this paper,
charge-conjugate modes are
{implied}
unless stated  otherwise.}}

The Belle detector is a large-solid-angle magnetic spectrometer that
consists of a silicon vertex detector (SVD),
a 50-layer central drift chamber (CDC),
an array of aerogel threshold Cherenkov counters (ACC), 
a barrel-like arrangement of
time-of-flight scintillation counters (TOF), 
and an electromagnetic calorimeter
comprised of 
CsI(Tl) {crystals (ECL), all located} inside
a superconducting solenoid coil
that provides a 1.5~T magnetic field.
An iron flux-return located {outside} the coil is instrumented to 
detect $K_{\rm{L}}^0$ mesons
and to identify muons (KLM).
The detector is described in detail elsewhere~\cite{Belle}.

{Leptons} are identified 
using likelihood ratios
calculated from 
the response of 
various subsystems of the detector.
{For electron identification,
the likelihood ratio is defined as 
{${\cal P}(e) = {\cal{L}}_e/({\cal{L}}_e+{\cal{L}}_x)$,}
where  ${\cal{L}}_e$ and ${\cal{L}}_x$ are the likelihoods 
for electron and non-electron, respectively,
determined using 
the ratio of the energy deposit in the ECL to 
the momentum measured in the SVD and CDC, 
the shower shape in the ECL, 
the matching between the position 
of {the} charged track trajectory and the cluster position in
the ECL,
the hit information from the {ACC,} 
and
the $dE/dx$ information in the CDC~\cite{EID}.
For muon  identification,
the likelihood ratio is defined as 
{${\cal P}(\mu) = {\cal{L}_\mu}/({\cal{L}}_{\mu}+{\cal{L}}_{\pi}+{\cal{L}}_{K})$,}
where  ${\cal{L}}_{\mu}$, ${\cal{L}}_\pi$  and ${\cal{L}}_K$ are the likelihoods 
for {the}
muon, pion and kaon {hypotheses,} respectively,
based on the matching quality and penetration depth of 
associated hits in the KLM~\cite{MUID}.}

In order to estimate the signal efficiency and 
to optimize the event selection, 
we use {Monte Carlo} (MC) samples.
The signal and the background events from generic $\tau^+\tau^-$ decays are 
 generated by KKMC/TAUOLA~\cite{KKMC}. 
In the signal MC, we generate $\tau^+\tau^-$, where 
a $\tau$ decays into three leptons 
{using {a 3-body-phase-space model~\cite{XX},}}
and the other $\tau$ decays 
generically.
Other {backgrounds,} including
$B\bar{B}$ and $e^+e^-\to q\bar{q}$ ($q=u,d,s,c$) events, Bhabha events, $e^+e^-\rightarrow\mu^+\mu^-$, 
and two-photon processes are generated by 
EvtGen~\cite{evtgen},
BHLUMI~\cite{BHLUMI}, 
KKMC and
{AAFH~\cite{AAFH}}, respectively. 
All kinematic variables are calculated in the laboratory frame
unless otherwise specified.
In particular,
variables
calculated in the $e^+e^-$ center-of-mass (CM) system
are indicated by the superscript ``CM''.

\section{Event Selection}

{We search for $\tau^+\tau^-$ events in which one $\tau$ 
(the {signal $\tau$}) decays into three leptons,
and the other $\tau$~(the {tag $\tau$}) decays 
into  one charged track, any number of additional 
photons and neutrinos.}
Candidate $\tau$-pair events are required to have 
four tracks with {zero} net charge.
{The following 
final states} 
are considered:
$e^-e^+e^-$,
$\mu^-\mu^+\mu^-$,
$e^-\mu^+\mu^-$,
$\mu^-e^+e^-$,
$\mu^-e^+\mu^-$, and
$e^-\mu^+e^-$.
{The event selection is optimized {mode-by-mode}
since the {backgrounds} are mode dependent.}

The event selection starts by reconstructing 
four charged tracks and any number of photons within the fiducial volume
defined by $-0.866 < \cos\theta < 0.956$,
{where $\theta$ is 
the polar angle {relative} to 
the direction opposite to 
that of 
the {incident} $e^+$ beam in 
{the} laboratory frame.}
The transverse momentum ($p_t$) of each charged track
and {the} energy of each photon ($E_{\gamma}$) 
are 
{required to satisfy} $p_t> $ 0.1 GeV/$c$ and $E_{\gamma}>0.1$ GeV,
respectively.
{For each charged track, 
the distance of the closest point with 
respect to the interaction point 
is required to be 
less than $\pm$0.5 cm in the transverse direction 
and less than 
$\pm$3.0 cm in the longitudinal direction.}

%
%
{Using the plane perpendicular to the CM
thrust axis~\cite{thrust},
which calculated from 
the observed tracks and photon candidates,
we separate the particles in an event
into two hemispheres.
These  are referred to as the signal and 
tag sides. 
The tag side contains one charged track}
while the signal side contains three charged tracks.
We require all charged tracks on the signal side 
{to be} 
identified as leptons.
The electron (muon)
identification criteria are ${\cal P}(e) > 0.9$ 
(${\cal P}(\mu) > 0.9$) 
{and}
momentum greater than 
0.3 GeV/$c$ (0.6 GeV/$c$).
The electron (muon) identification
{efficiency
is} 91\% (85\%) 
while 
{the probability to misidentify {a} pion 
as 
{an} {electron} ({a} muon)}
is below 0.5\% (2\%).

%
%
To ensure that the missing particles are neutrinos rather
than photons or charged particles that pass  outside the detector acceptance,
we impose requirements on the missing 
momentum $\vec{p}_{\rm miss}$,
which 
is
calculated by subtracting the
vector sum of the momenta of all tracks and photons
from the sum of the $e^+$ and $e^-$ beam momenta.
We require that the magnitude of $\vec{p}_{\rm miss}$
be  greater than 0.4 GeV/$c$,
and {that} its direction point into the fiducial volume of the
detector.

%
%
To reject $q\bar{q}$ background,
we require that the magnitude of thrust ($T$)
be 0.90 $< T  <$ 0.97 
for all modes except for the $\tau^-\to e^-e^+e^-$ {mode,} 
{for 
{which it is} 0.90 $ < T  <$ 0.96.}
We also require $5.29$ GeV $< E^{\mbox{\rm{\tiny{CM}}}}_{\rm{vis}} < 9.5$ GeV, 
where $E^{\mbox{\rm{\tiny{CM}}}}_{\rm{vis}}$ 
is the total visible energy in the CM system, defined as 
the sum of the energies of {the} three leptons,
the charged track on the tag side (with a pion mass hypothesis{),}
and all photon candidates.

%
%
Since neutrinos are 
emitted only on the tag side,
the direction of
$\vec{p}_{\rm miss}$
should lie within the tag side of the event.
The cosine of the
opening angle between
$\vec{p}_{\rm miss}$
and the charged track on the tag side 
{in the CM system,}
$\cos \theta^{\mbox{\rm \tiny CM}}_{\rm tag-miss}$, 
{is 
required to lie
in the range 
$0.0<\cos \theta^{\mbox{\rm \tiny CM}}_{\rm tag-miss}<0.98$.
{This upper} limit 
reduces background
from
Bhabha, $\mu^+\mu^-$ and two-photon background events,
as 
radiated gammas 
{from the tag-side track 
result in missing momentum
if they overlap with
the ECL clusters of} 
{the} tag-side track~\cite{cite:tau_egamma}.}}
The reconstructed mass on
the tag side using a charged track (with a pion mass hypothesis) and photons,
$m_{\rm tag}$, 
is required to be less than 1.777 GeV/$c^2$.
As shown in Fig.~\ref{fig:cut_fig},
reasonable agreement between data and 
{the} background expectation 
{from MC simulation}
is obtained in  the distributions of 
$\cos \theta^{\mbox{\rm \tiny CM}}_{\rm tag-miss}$  
and $m_{\rm tag}$.

%
%

\begin{figure}
 \resizebox{0.37\textwidth}{0.37\textwidth}{\includegraphics
{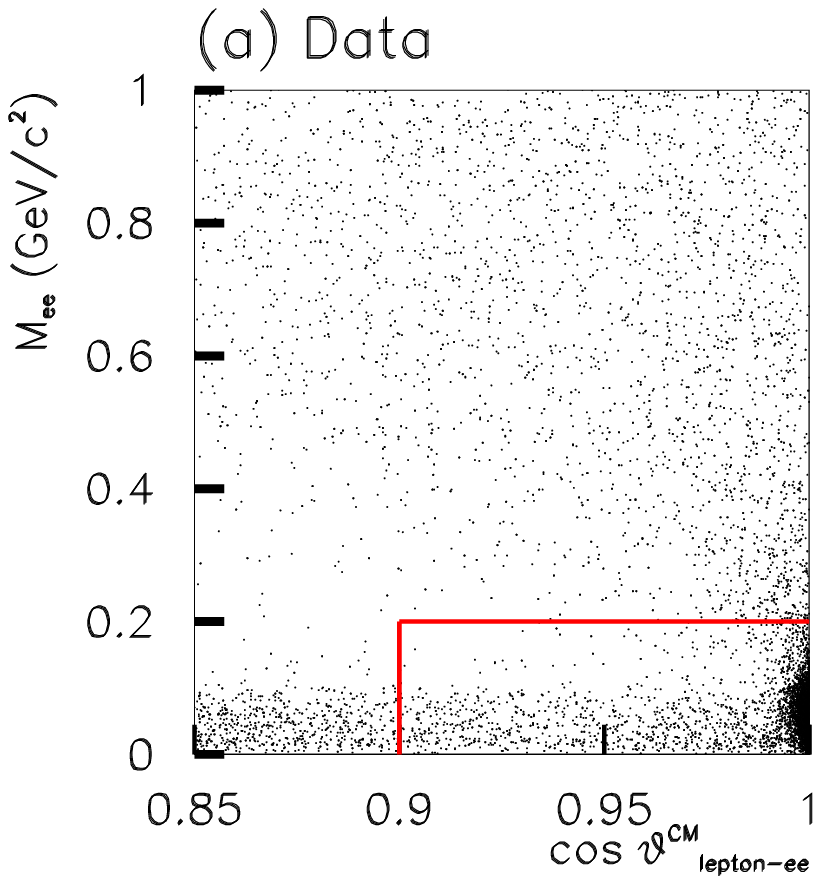}}
\hspace*{-1.2cm}
 \resizebox{0.37\textwidth}{0.37\textwidth}{\includegraphics
{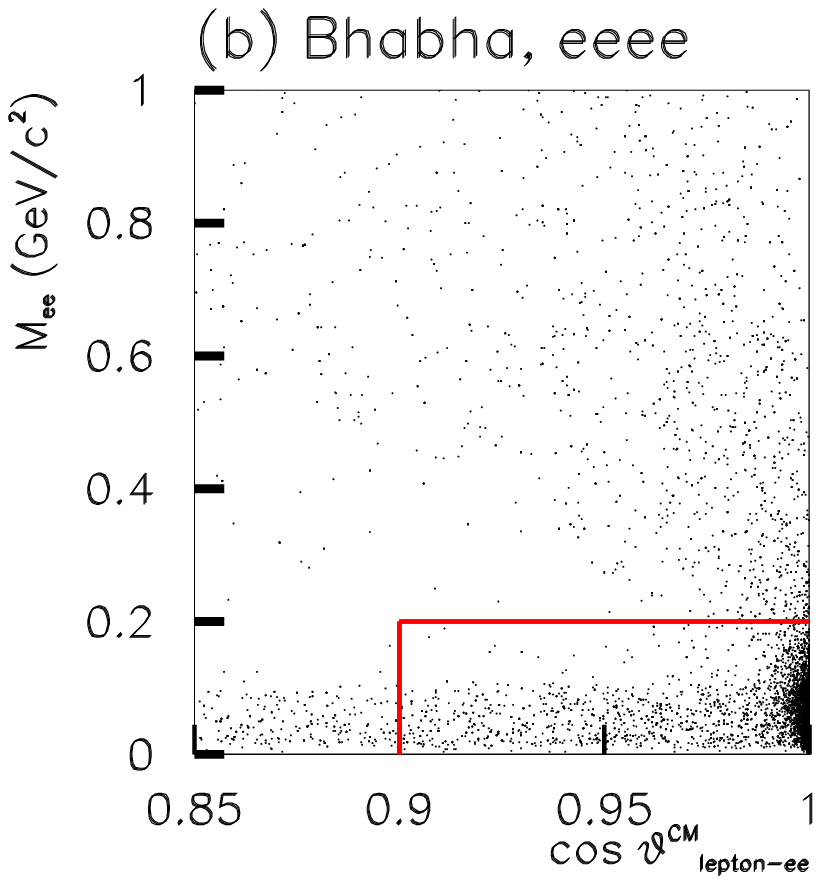}}
\hspace*{-1.2cm}
 \resizebox{0.37\textwidth}{0.37\textwidth}{\includegraphics
{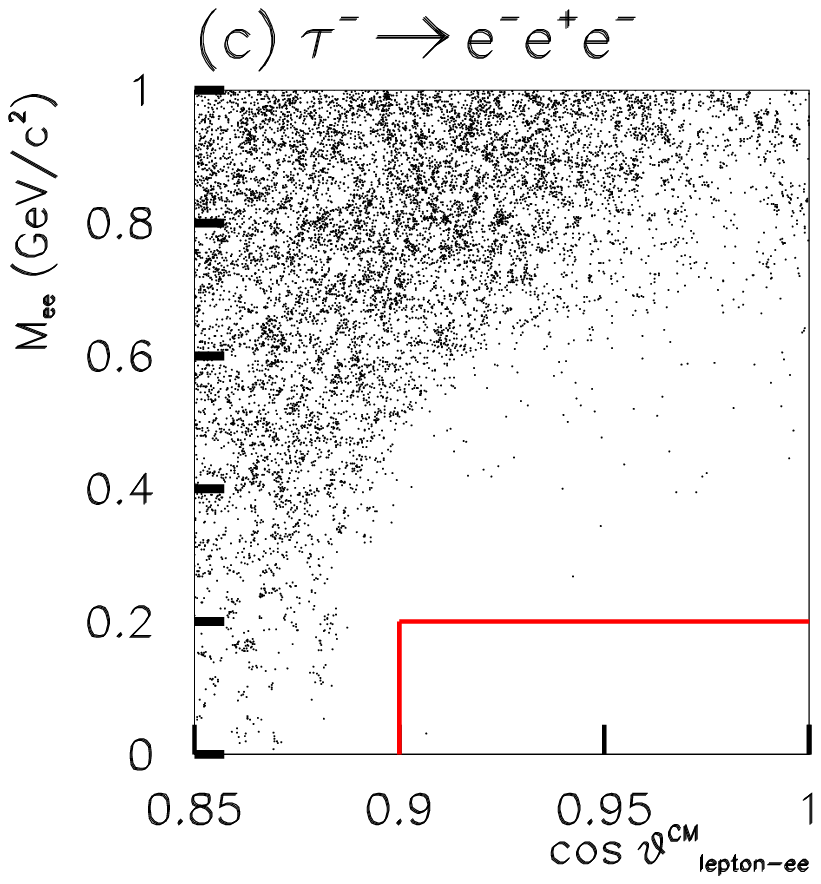}}
 \vspace*{-0.5cm}
\caption{
Scatter-plots
of the
{reconstructed  invariant mass} 
of the $e^+e^-$ pair
($M_{ee}$) vs.  cosine of the  opening angle between
the direction of the $e^+e^-$ pair and the 
other electron 
($\cos \theta^{\rm CM}_{\rm{lepton}-ee}$)
for
(a) data, (b) Bhabha and $eeee$,
(c) signal MC ($\tau^-\to e^-e^+e^-$)
}
\label{fig:convsersion}
\end{figure}

Conversions ($\gamma\to e^+e^-$) 
are a large background for the 
{$\tau^-\-\to e^-e^+e^-$} and $\mu^-e^+e^-$ modes.
{For these modes,
if the invariant mass of the $e^+e^-$ pair
($M_{ee}$) is less than 0.2 GeV/$c^2$,
we require that 
the cosine of the opening angle in the {CM} system
between 
the momentum 
of the $e^+e^-$ pair 
and the momentum of the other lepton
($\cos \theta^{\rm CM}_{\rm{lepton}-ee}$)
be less than 0.90.}
As shown in Fig.~\ref{fig:convsersion} 
{for the $\tau^-\to e^-e^+e^-$ mode 
(two entries {for} each event),} 
the signal efficiency is not affected by this cut, 
{while the large background from 
conversions is substantially reduced.}

%
%
For the $\tau^-\to e^-e^+e^-$ and
$\tau^-\to e^-\mu^+\mu^-$ modes,
the charged track on the tag side is required not to be an electron 
{by applying 
${\cal P}(e)<0.1$;
this reduces 
large backgrounds 
from two-photon and Bhabha {processes}
that remain.}
Furthermore,
we reject the event if the charged track on the tag side 
{traverses the 
gap between the barrel and the endcap} of
the ECL.
To reduce 
{Bhabha and $\mu^+\mu^-$
{backgrounds},}
we require that the momentum in the CM system of 
the charged track on the tag side be less than 4.5 {GeV/$c$}
for the $\tau^-\to e^-e^+e^-$ and $\tau^-\to\mu^-e^+e^-$ modes.

\begin{figure}
\begin{center}
 \resizebox{0.65\textwidth}{0.65\textwidth}{\includegraphics
 {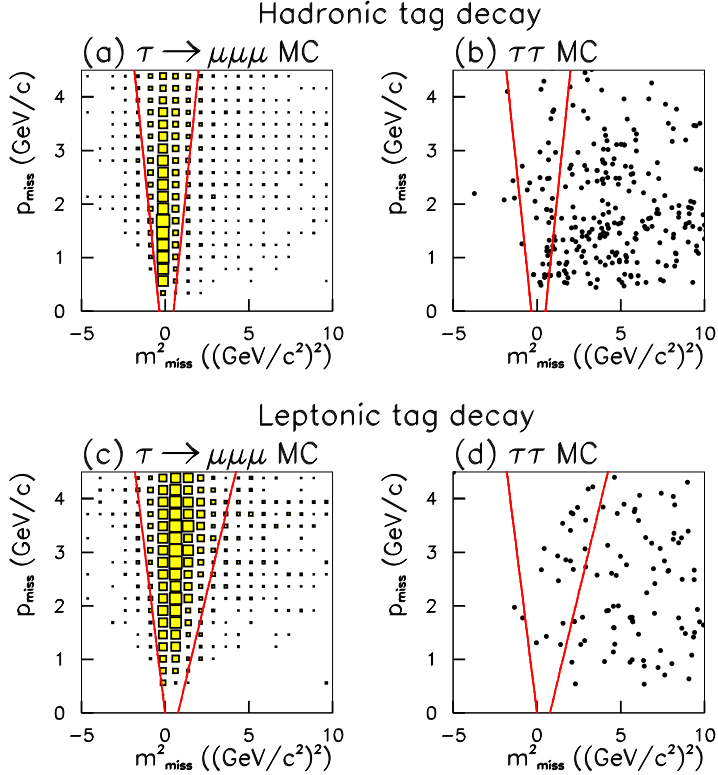}}
 \vspace*{-0.5cm}
 \caption{
Scatter-plots of
{$p_{\rm miss}$ 
{vs.} 
$m_{\rm miss}^2$:
(a) and (b)
show
the signal MC 
($\tau^-\to\mu^-\mu^+\mu^-$)
and 
the generic $\tau^+\tau^-$ MC 
distributions,
respectively,
for the hadronic tags
while (c) and
(d) show
the same distributions
for the {leptonic tags.}}
Selected regions are indicated by lines.}
\label{fig:pmiss_vs_mmiss2}
\end{center}
\end{figure}

%
%
Finally, 
to suppress  backgrounds from generic 
$\tau^+\tau^-$ and $q\bar{q}$ events, 
we apply a selection based on the magnitude of the missing momentum ${p}_{\rm{miss}}$ 
and {the}
missing mass squared $m^2_{\rm{miss}}$ for all modes 
except for 
$\tau^-\to e^+\mu^-\mu^-$ and $\mu^+ e^-e^-$.
We do not apply
this cut 
for the latter two modes
since  {backgrounds in these cases} are 
much {smaller.}
{We apply different selection criteria according to the lepton
identification of the charged track {on the tag side, as} 
the number of
emitted neutrinos is two if the track is an electron or muon 
(leptonic tag) while it is one if the track is a hadron (hadronic tag).}
The selection criteria are listed in Table~\ref{tbl:misscut};  
the distributions of $m^2_{\rm{miss}}$ and $p_{\rm{miss}}$ for 
hadronic and leptonic decays are shown in Fig.~\ref{fig:pmiss_vs_mmiss2}. 
\begin{table}
\begin{center}
\caption{
The selection criteria for 
the missing momentum ($p_{\rm{miss}}$) and
missing mass squared ($m^2_{\rm miss}$) 
for each mode,
The units for $p_{{\rm miss}}$ 
and  $m^2_{\rm miss}$ 
{are}  GeV/$c$ and $({\rm{GeV}}/c^2)^2$, respectively.}
\label{tbl:misscut}
\begin{tabular}{c|c|c} \hline\hline
Mode & Hadronic tag mode & Leptonic tag mode \\ \hline
$\tau^-\to\mu^-\mu^+\mu^-$ & $p_{\rm miss} > -3.0~m^2_{\rm miss}-1.0$ &
 $p_{\rm miss} > -2.5~m^2_{\rm miss}$ \\
$\tau^-\to \mu^- e^+ e^-$ &  $p_{\rm miss} > 3.0~m^2_{\rm miss}-1.5$  &
 $p_{\rm miss} > 1.3~m^2_{\rm miss}-1$ \\
$\tau^-\to e^- \mu^+ \mu^-$ & & \\ \hline
$\tau^-\to e^-e^+ e^-$ & $p_{\rm miss} > -3.0~m^2_{\rm miss}-1.0$ &
 $p_{\rm miss} > -2.5~m^2_{\rm miss}$ \\
 &  $p_{\rm miss} > 4.2~m^2_{\rm miss}-1.5$  &
 $p_{\rm miss} > 2.0~m^2_{\rm miss}-1$ \\ \hline
$\tau^-\to e^+ \mu^- \mu^-$ & {Not applied}  & 
{Not applied} \\
$\tau^-\to \mu^+ e^- e^-$ &    & \\ \hline\hline
\end{tabular}
\end{center}
\end{table}

\section{Signal and Background Estimation}

The signal candidates are examined in the 
two-dimensional {plot} of the $\ell^-\ell^+\ell^-$ invariant
mass~($M_{\rm {3\ell}}$) {versus} 
the difference of their energy from the 
beam energy in the CM system~($\Delta E$).
A signal event should have $M_{\rm {3\ell}}$
close to the $\tau$-lepton mass and
$\Delta E$ close to zero.
For all modes,
the $M_{\rm {3\ell}}$ and $\Delta E$  resolutions are parameterized
from fits to the signal MC {distributions,}  
with  an asymmetric Gaussian function that takes into account 
{initial-state} radiation.
The resolutions {in}
$M_{\rm 3\ell}$ and $\Delta E$ for each mode are summarized 
in Table~\ref{tbl:reso_del_e_m}.
\begin{table}
\begin{center}
\caption{Summary {of} $M_{\rm 3\ell}$  and 
$\Delta E$ 
{resolutions}.
The $\sigma^{\rm high}$ ($\sigma^{\rm low}$) 
means the standard deviation 
{on the} higher (lower) side of the peak.}
\label{tbl:reso_del_e_m}
\begin{tabular}{c|cccc} \hline\hline 
Mode
& $\sigma^{\rm{high}}_{M_{\rm{3\ell}}}$ (MeV/$c^2$)  
& $\sigma^{\rm{low}}_{M_{\rm{3\ell}}}$ (MeV/$c^2$)
& $\sigma^{\rm{high}}_{\Delta E}$ (MeV)      
&  $\sigma^{\rm{low}}_{\Delta E}$ (MeV)
 \\ \hline
$\tau^-\to\mu^-\mu^+\mu^-$
& 4.8  &  5.4 & 12.5 & 15.7 \\
$\tau^-\to e^- e^+ e^-$
& 5.1 & 7.8 & 13.4 & 25.1 \\
$\tau^-\to e^-\mu^+\mu^-$
& 5.1  &  5.6 & 12.1 & 19.6  \\
$\tau^-\to\mu^- e^+ e^-$
 & 5.0  &  6.6 & 13.4 & 21.3  \\
$\tau^-\to e^+\mu^-\mu^-$
& 5.0  &  6.0 & 13.3 & 19.9  \\
 $\tau^-\to\mu^+ e^- e^-$
 & 5.4  &  6.7 & 13.8 & 23.0  \\  \hline\hline
\end{tabular}
\end{center}
\end{table}

\begin{figure}
\begin{center}
\resizebox{0.4\textwidth}{0.4\textwidth}{\includegraphics
{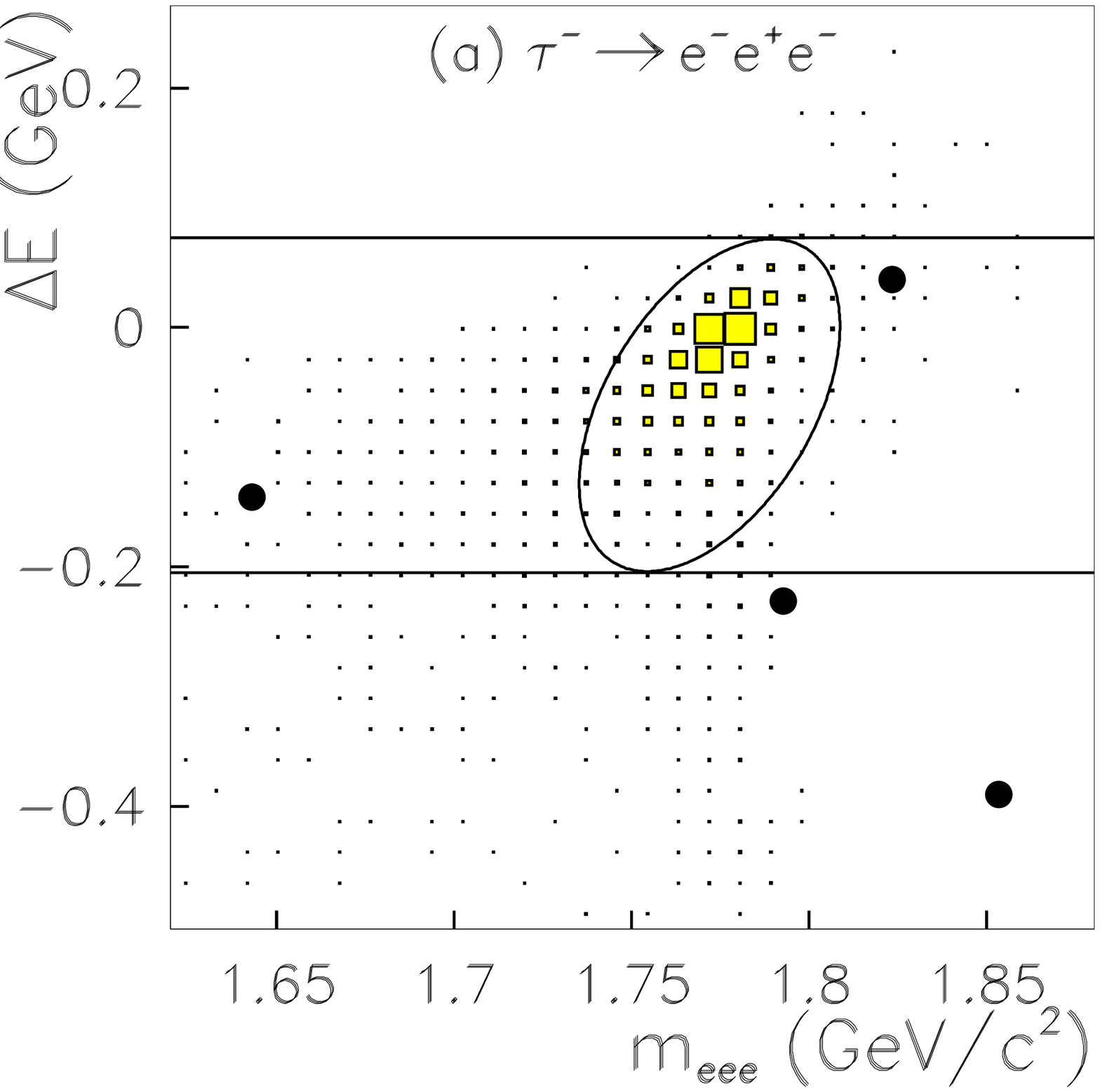}}
\resizebox{0.4\textwidth}{0.4\textwidth}{\includegraphics
{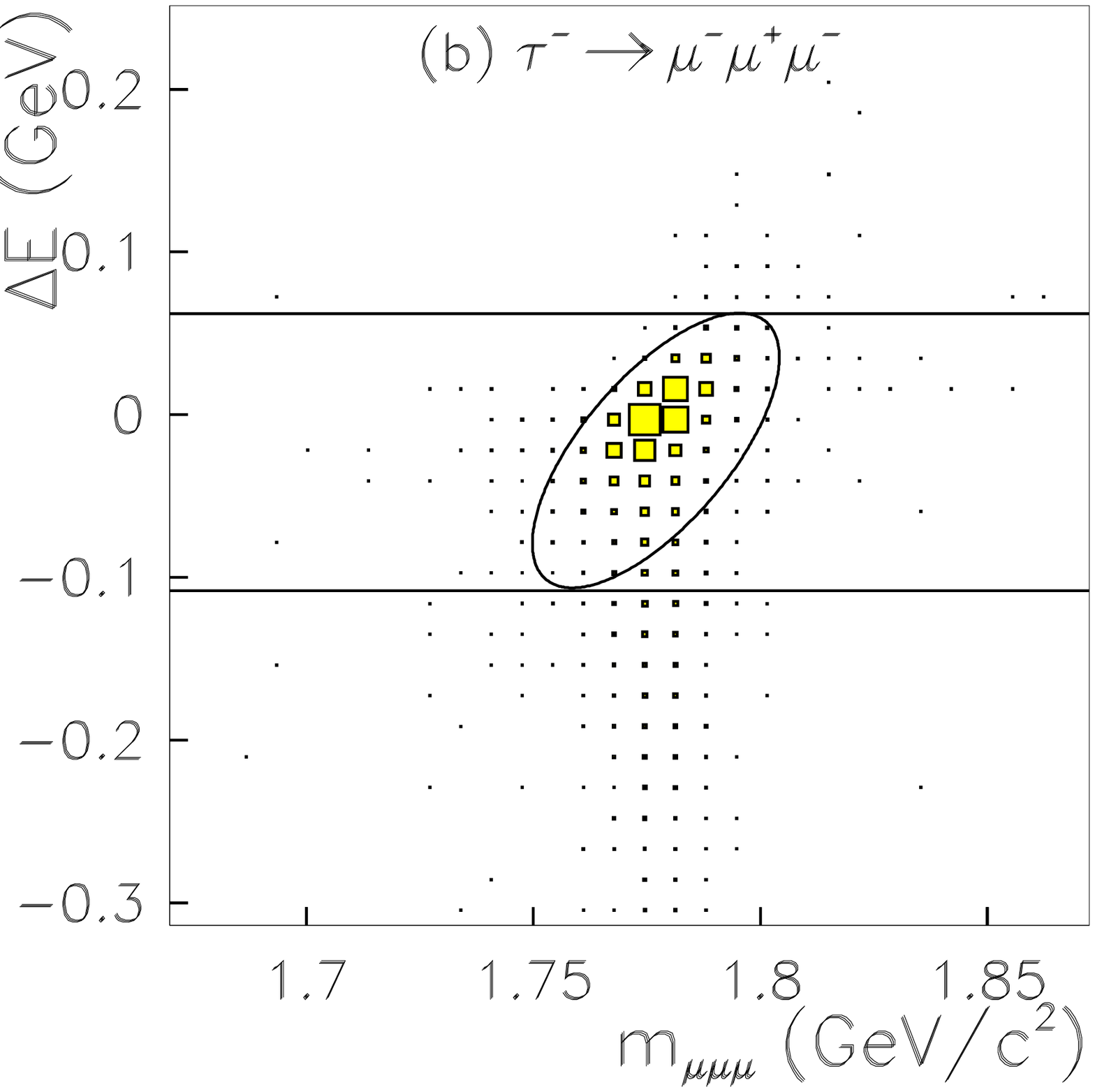}}\\
\vspace*{-0.5cm}
\resizebox{0.4\textwidth}{0.4\textwidth}{\includegraphics
{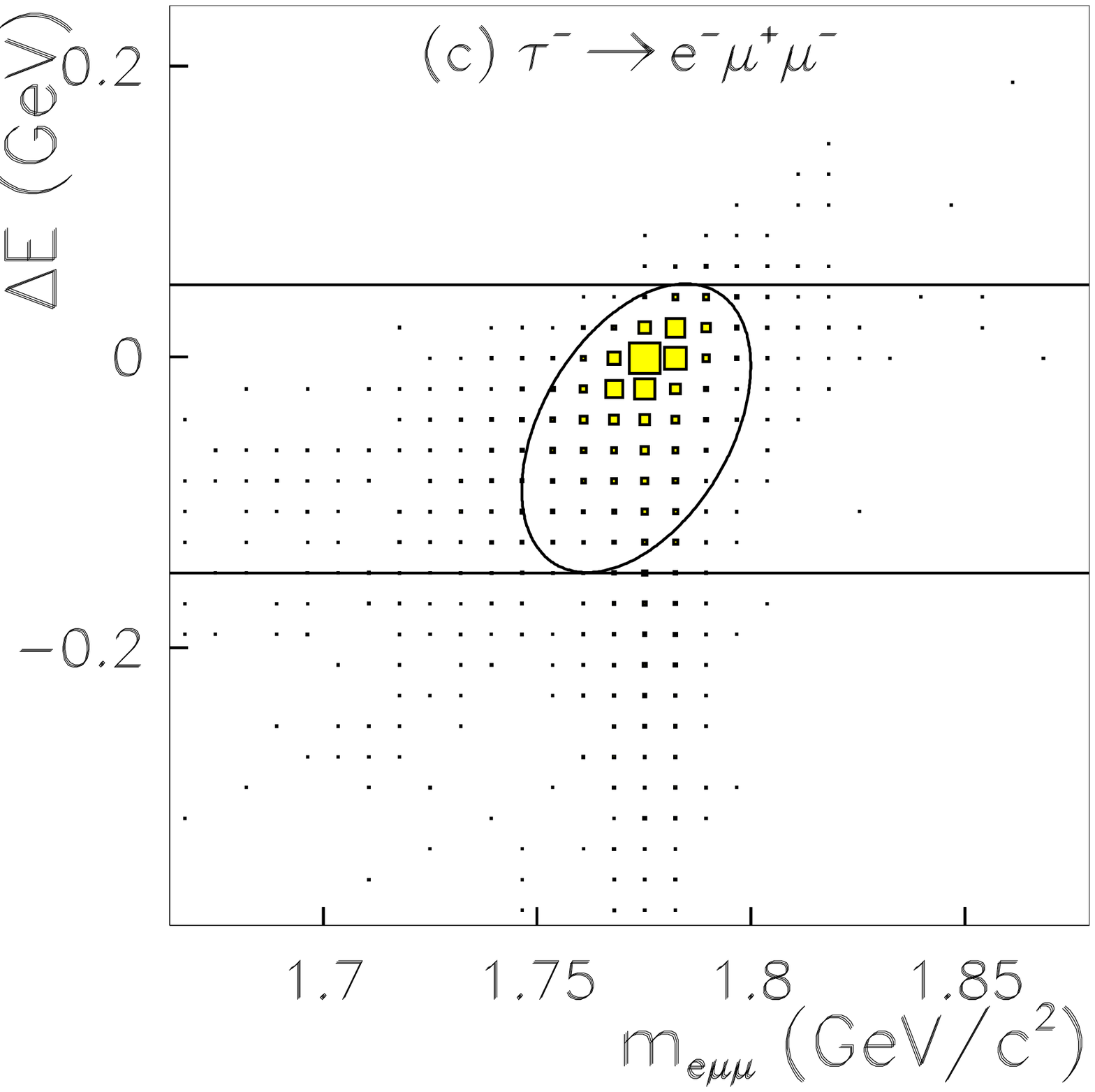}}
\resizebox{0.4\textwidth}{0.4\textwidth}{\includegraphics
{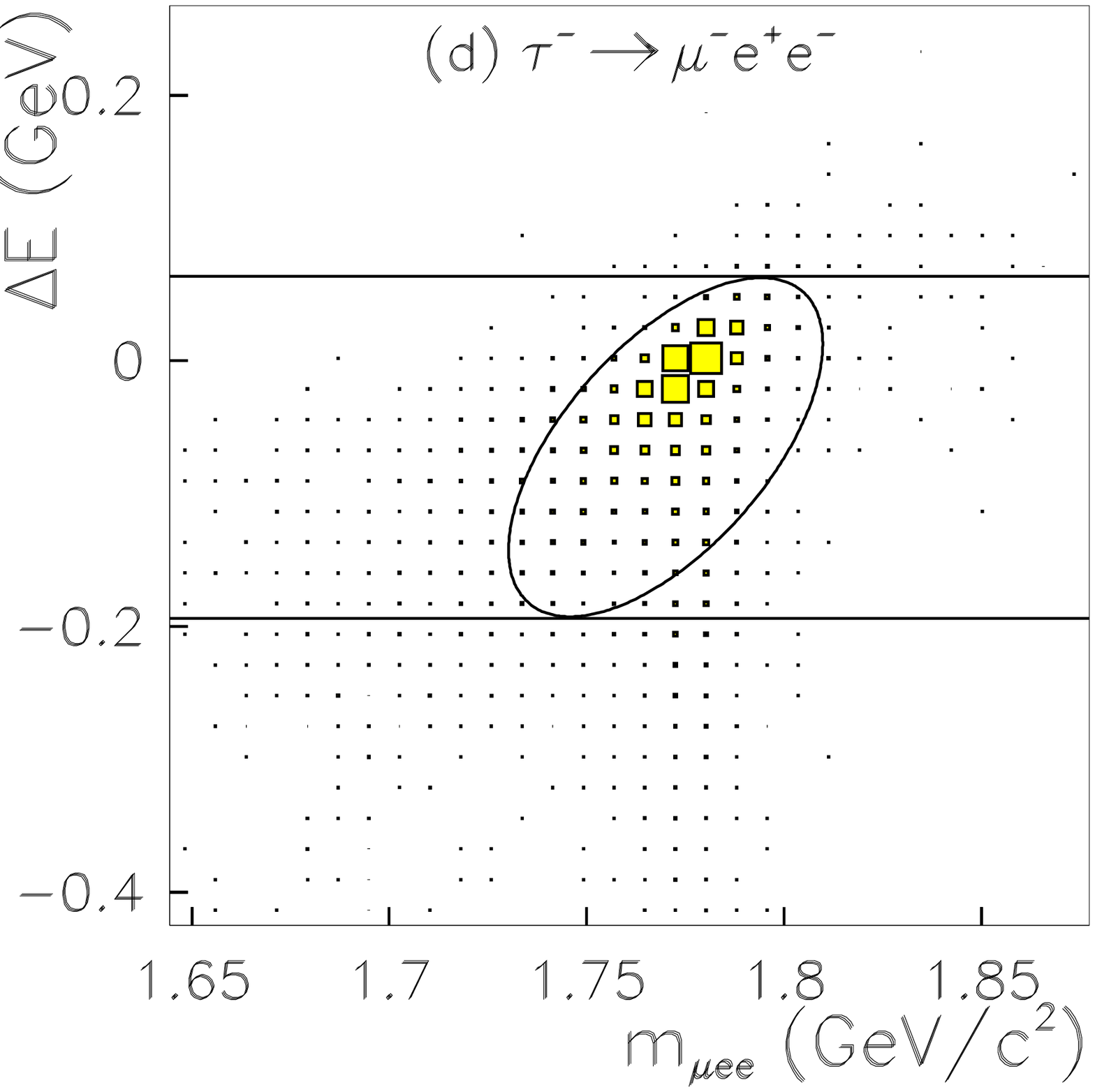}}\\
\vspace*{-0.5cm}
\resizebox{0.4\textwidth}{0.4\textwidth}{\includegraphics
{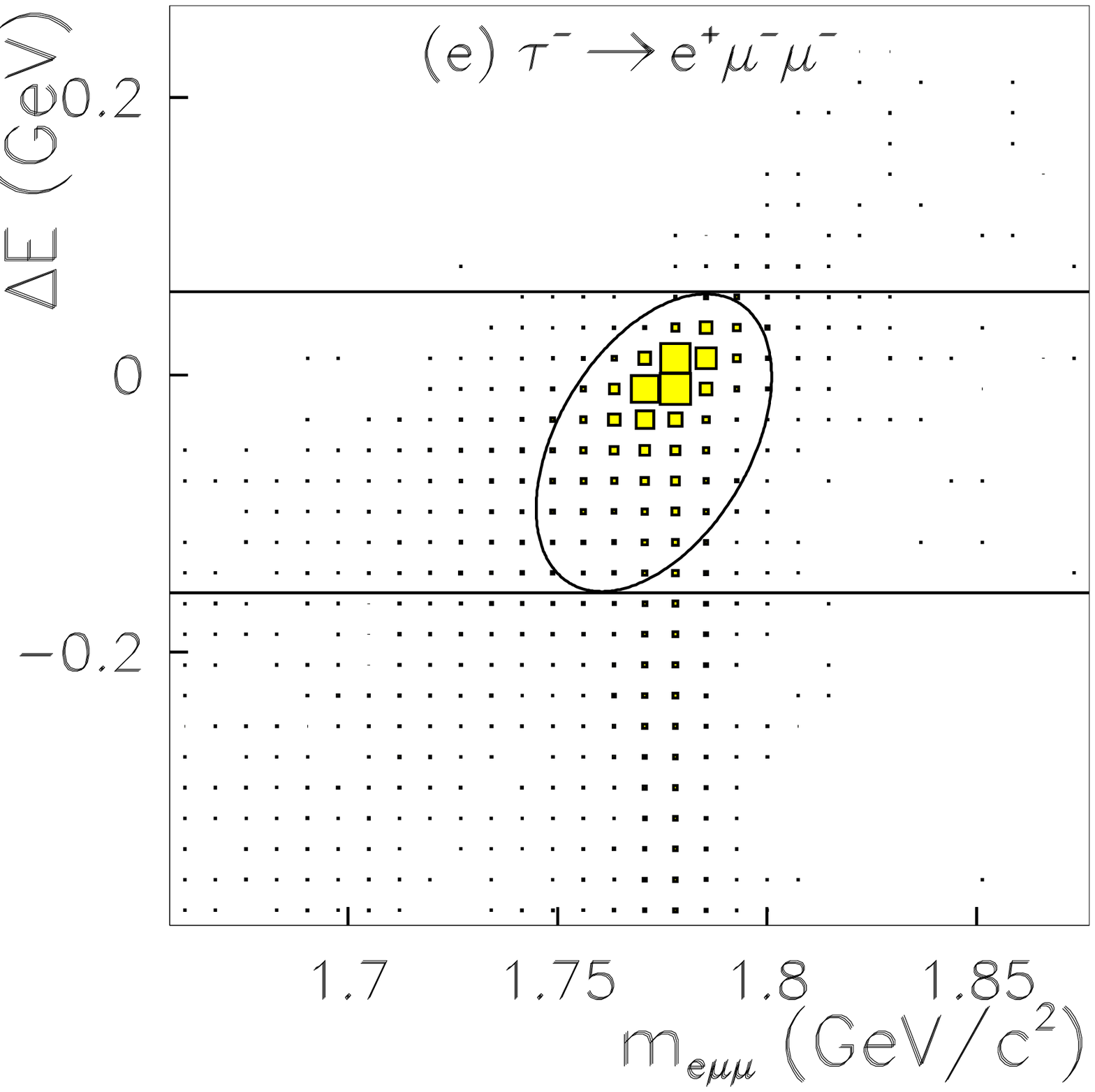}}
\resizebox{0.4\textwidth}{0.4\textwidth}{\includegraphics
{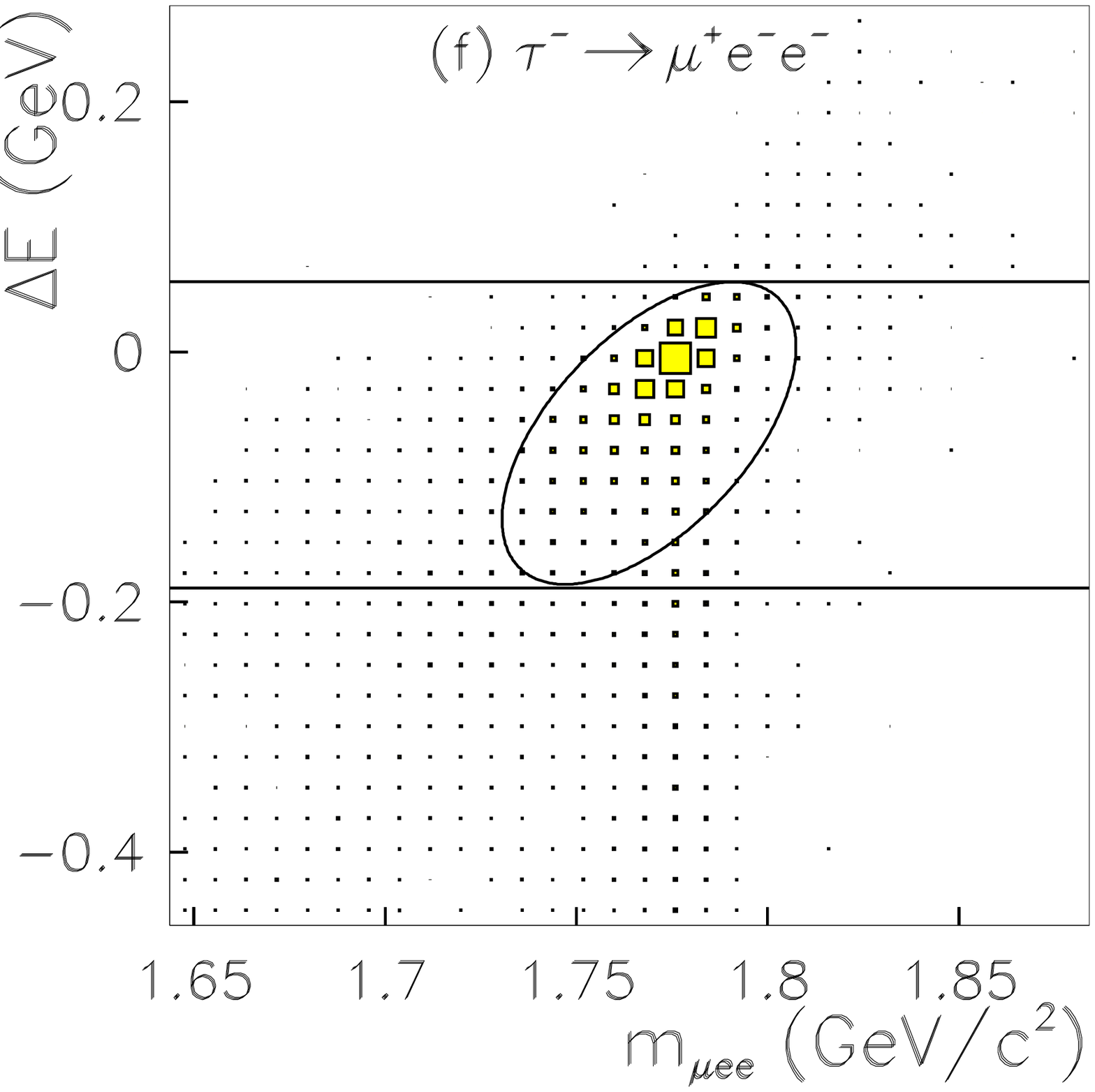}}
\caption{
Scatter-plots in the
$M_{3\ell}$ -- $\Delta{E}$ plane:
(a), (b), (c), (d), (e) and (f) correspond to
the $\pm 20 \sigma$ area for
the
$\tau^-\rightarrow e^-e^+e^-$,
$\tau^-\rightarrow\mu^-\mu^+\mu^-$,
$\tau^-\rightarrow e^-\mu^+\mu^-$,
$\tau^-\rightarrow\mu^- e^+e^-$,
$\tau^-\rightarrow e^+\mu^-\mu^-$ and
$\tau^-\rightarrow \mu^+e^-e^-$
modes, respectively.
The data are indicated by the solid circles.
The filled boxes show the MC signal distribution
with arbitrary normalization.
The elliptical signal 
{regions} 
shown by a solid curve 
are used for evaluating the signal yield.
The region between the horizontal solid lines excluding
the signal region is
used to estimate the expected background in the elliptical region. 
}
\label{fig:openbox}
\end{center}
\end{figure}

{{To evaluate the branching {fractions,}
we use  {elliptical signal regions}
{that} {contain} 90\%
of the MC signal {events satisfying} all {selection criteria.}
We blind the data in the signal region
{until all selection criteria are finalized}
so as not to bias our choice of selection criteria. }}
Figure~\ref{fig:openbox} shows scatter-plots
for the data and the signal MC distributed over $\pm 20\sigma$
in the $M_{\rm{3\ell}}-\Delta E$ plane.
No events are observed 
outside the signal region 
for any modes except 
for 
{$\tau^-\to e^-e^+e^-$,} in which four events are found.
{These remaining events all
have $e^+e^-$ invariant masses below 0.1 GeV/$c^2$,
{and we} nominally attribute them
to  Bhabha  electron or $\tau^-\to e^-\nu_{\tau}\bar{\nu_{e}}$ 
processes accompanied by a gamma conversion.}
{The final estimate of 
background} 
is based on the data 
with looser selection criteria
{for} 
particle identification and event selection
in the $M_{\rm{3\ell}}$ 
{sideband region. 
The sideband region is} defined as the box 
inside
{the} horizontal lines {but}
excluding the signal region,
as shown by the lines in Fig.~\ref{fig:openbox}.
Assuming that the background distribution is uniform in the sideband region, 
the number of background events in the signal box is estimated by 
interpolating the number of observed events in the sideband {region} 
into the signal {region.}
The signal efficiency and 
the number of expected background 
events for each mode 
are summarized in Table~\ref{tbl:eff}.
{After estimating the background,  
we 
{unblind
and 
{find no candidate events} 
for any of the modes.}}

We estimate systematic uncertainties
{due to  lepton identification, 
charged track finding,  MC statistics, and  integrated luminosity.} 
The uncertainty due to the trigger efficiency is negligible compared with the other uncertainties.
{The uncertainties due to lepton {identification} 
{are} 
2.2\% per  electron and 2.0\% per  muon.}
The uncertainty due to the charged track finding is estimated to be 1.0\% per charged track.
The uncertainty due to the $e$-veto on the tag side applied for 
{the} 
$\tau^-\to e^-e^+e^-$ and $\tau^-\to e^-\mu^+\mu^-$ modes is estimated to be the same as  
the uncertainty due to the electron identification.
The uncertainties due to MC statistics and luminosity
are estimated to be 
{(0.5 - 0.9)\%} and 1.4\%, respectively.
All these uncertainties are added in quadrature, 
and the total systematic uncertainty for each mode is listed in Table~\ref{tbl:eff}.
\begin{table}
\begin{center}
\caption{ The signal efficiency($\varepsilon$), 
the number of the expected background {events}  ($N_{\rm BG}$)
estimated from the  sideband data, 
{total} 
systematic uncertainty  ($\sigma_{\rm syst}$),
{the} number of the observed events 
in the signal region ($N_{\rm obs}$), 
90\% C.L. upper limit on the number of signal events including 
systematic uncertainties~($s_{90}$) 
and 90\% C.L. upper limit on the branching 
fraction~($\cal{B}$)
for each individual mode. }
\label{tbl:eff}
\begin{tabular}{c|cccccc}\hline \hline
Mode &  $\varepsilon$~{(\%)} & 
$N_{\rm BG}$  & $\sigma_{\rm syst}$~{(\%)}
& $N_{\rm obs}$ & $s_{90}$ & 
${\cal{B}}(\times10^{-8})$ \\ \hline
$\tau^-\to e^-e^+e^-$ &  6.00 & 
 0.40$\pm$0.30 & 9.8 &
 0 & 2.10 & 3.6 \\ 
$\tau^-\to\mu^-\mu^+\mu^-$ & 7.64 & 0.07$\pm{0.05}$ & 
 7.4 &
0 &  2.41 & 3.2 \\
$\tau^-\to e^-\mu^+\mu^-$ &  6.08 & 0.05$\pm{0.03}$ 
  & 9.5 &
 0 & 2.44 & 4.1\\
$\tau^-\to \mu^-e^+e^-$ &  9.29 & 0.04$\pm{0.04}$
  & 7.8 &
0 & 2.43  & 2.7\\ 
$\tau^-\to e^+\mu^-\mu^-$ &  10.8 & 0.02$\pm{0.02}$
 & 7.6 & 
0 &  2.44& 2.3\\
$\tau^-\to \mu^+e^-e^-$ &  12.5 &
0.01$\pm{0.01}$
 & 7.7 & 
0 & 2.46  & 2.0\\ 
\hline\hline
\end{tabular}
\end{center}
\end{table}

\section{Upper Limits on the branching fractions}

We set upper limits on the branching fractions 
of $\tau^-\to\ell^-\ell^+\ell^-$
based on the Feldman-Cousins method~\cite{cite:FC}.
The 90\% C.L. upper limit on the number of the signal events 
and including  a systematic uncertainty~($s_{90}$) 
is obtained 
{using} the POLE program without conditioning \cite{pole}
{based on}
the number of expected {background events}, observed data
and the systematic uncertainty.
The upper limit on the branching fraction ($\cal{B}$) is then given by
\begin{equation}
{{\cal{B}}(\tau^-\to\ell^-\ell^+\ell^-) <
\displaystyle{\frac{s_{90}}{2N_{\tau\tau}\varepsilon{}}},}
\end{equation}
where $N_{\tau\tau}$ is the number of $\tau^+\tau^-$pairs, and 
$\varepsilon$ is the signal efficiency.
{The value {$N_{\tau\tau} =  492\times 10^6$}} is obtained 
from 
{the} integrated luminosity times 
the cross section of {$\tau$-pair production,} which 
is calculated 
{in the updated version of 
KKMC~\cite{tautaucs} to be 
$\sigma_{\tau\tau} = 0.919 \pm 0.003$ nb.}
The 90\% C.L. upper limits on the branching fractions 
${\cal{B}}(\tau^-\rightarrow \ell^- \ell^+\ell^-)$  are in the range
between 
{$2.0 \times 10^{-8}$ and $4.1 \times 10^{-8}$} 
and are summarized in Table~\ref{tbl:eff}.
These results improve  {upon} our 
previously published upper limits~\cite{cite:3l_belle} 
{by factors of {4.9 to 10.}}
{They are also more stringent upper limits 
than the recent BaBar results~\cite{cite:3l_babar2},
except for the $\tau^-\to e^-\mu^+\mu^-$ {mode,}
for which the limit is similar.}

\section{Summary}
We have searched for {lepton-flavor-violating} $\tau$ decays 
into three leptons using 535 fb$^{-1}$ of data.
No events 
are observed and
we set  90\% C.L. upper limits 
{on the branching fractions:} 
${\cal{B}}(\tau^-\rightarrow e^-e^+e^-) < 3.6\times 10^{-8}$, 
${\cal{B}}(\tau^-\rightarrow \mu^-\mu^+\mu^-) < 3.2\times 10^{-8}$, 
${\cal{B}}(\tau^-\rightarrow e^-\mu^+\mu^-) < 4.1\times 10^{-8}$, 
${\cal{B}}(\tau^-\rightarrow \mu^-e^+e^-) < 2.7\times 10^{-8}$, 
${\cal{B}}(\tau^-\rightarrow e^+\mu^-\mu^-) < 2.3\times 10^{-8}$
and  
${\cal{B}}(\tau^-\rightarrow \mu^+e^-e^-) < 2.0\times 10^{-8}$.
These results improve {upon} {our} 
previously published upper limits
by factors {of} 4.9 to 10.
These more stringent upper limits can be used
to constrain the space of parameters in various models of new physics.

\section*{Acknowledgments}

The authors are grateful to A.~Buras and Th.~Mannel for fruitful 
discussions.
We thank the KEKB group for the excellent operation of the
accelerator, the KEK cryogenics group for the efficient
operation of the solenoid, and the KEK computer group and
the National Institute of Informatics for valuable computing
and Super-SINET network support. We acknowledge support from
the Ministry of Education, Culture, Sports, Science, and
Technology of Japan and the Japan Society for the Promotion
of Science; the Australian Research Council and the
Australian Department of Education, Science and Training;
the National Science Foundation of China and the Knowledge
Innovation Program of the Chinese Academy of Sciences under
contract No.~10575109 and IHEP-U-503; the Department of
Science and Technology of India; 
the BK21 program of the Ministry of Education of Korea, 
the CHEP SRC program and Basic Research program 
(grant No.~R01-2005-000-10089-0) of the Korea Science and
Engineering Foundation, and the Pure Basic Research Group 
program of the Korea Research Foundation; 
the Polish State Committee for Scientific Research; 
the Ministry of Education and Science of the Russian
Federation and the Russian Federal Agency for Atomic Energy;
the Slovenian Research Agency;  the Swiss
National Science Foundation; the National Science Council
and the Ministry of Education of Taiwan; and the U.S.\
Department of Energy.

\end{document}